\documentclass[prd,showpacs,twocolumn]{revtex4}
\usepackage{chay,epsfig}
\begin{document}
\title{Factorization of $B$ decays into two light mesons in soft-collinear
  effective theory} 
\author{Junegone Chay}
\author{Chul Kim}
\affiliation{Department of Physics, Korea University, Seoul 136-701,
Korea} 
\preprint{KUPT-03-01}
\begin{abstract}
We show that the amplitudes for $B$ decays into two
light mesons at leading order in soft-collinear effective theory are
factorized to all orders in $\alpha_s$. We construct gauge-invariant
four-quark operators by employing the effective theories
$\mathrm{SCET}_{\mathrm{I}}$ for $\sqrt{m_b \Lambda} <\mu<m_b$, and
$\mathrm{SCET}_{\mathrm{II}}$ for $\mu <\sqrt{m_b \Lambda}$ with
$\Lambda \sim 0.5$ GeV. These operators do not allow gluon exchanges
between different sectors of the current and the matrix elements are
reduced to the products of current matrix elements.
The spectator interactions are also factorized at
leading order in SCET and to all orders in $\alpha_s$.

\end{abstract}

\pacs{13.25.Hw, 11.10.Hi, 12.38.Bx, 11.40.-q}

\maketitle
Nonleptonic $B$ decays have been the subject of intense interest and
precise experimental observation of nonleptonic $B$ decays makes it
urgent to give firm theoretical prediction including CP
violation. The theoretical treatment of nonperturbative effects from
the strong interaction in nonleptonic $B$ decays is a deep problem for
a long time, and the soft-collinear effective theory casts a new
perspective.   

Recently the soft-collinear effective theory (SCET) has
been formulated \cite{scet,beneke1}, and it is applied to various cases in
which energetic light quarks appear. For example, the factorization in
$B\rightarrow D\pi$ decays \cite{bauer2}, the heavy-to-light form
factor \cite{form,chay} and the pion form factor \cite{bauer3} were
considered. We apply SCET to  
$B$ decays into two light mesons such as $B \rightarrow
\pi \pi$  to show factorization properties. In this problem, we 
have to consider the operators consisting of the  collinear fields
moving in the opposite direction and a heavy quark.

We employ two effective theories, namely, $\mathrm{SCET}_{\mathrm{I}}$
for $\sqrt{m_b \Lambda} <\mu <m_b$ with $\Lambda \sim 0.5$ GeV, and
$\mathrm{SCET}_{\mathrm{II}}$ for $\mu< \sqrt{m_b \Lambda}$. 
In SCET, there are collinear, soft and ultrasoft (usoft) fields which
are classified by the scaling of their momenta. The collinear momentum
scales as $p_n^{\mu} = 
(n\cdot p_n, \overline{n} \cdot p_n, p_n^{\perp}) \sim m_b (\lambda^2,
1,\lambda)$ for the colliear particles in the $n^{\mu}$ 
direction, and $p_{\overline{n}}^{\mu} \sim m_b (1,\lambda^2,\lambda)$
in the $\overline{n}^{\mu}$ direction, where $n^2=\overline{n}^2=0$,
$n\cdot \overline{n}=2$, and $\lambda \sim \sqrt{\Lambda/m_b} \ll 1$
is the small expansion parameter. Soft momentum scales  
as $p_s^{\mu} \sim m_b (\lambda,\lambda,\lambda)$, but soft particles
do not interact with collinear particles due to momentum conservation.
And the usoft momentum scales as $p_{\mathrm{us}}^{\mu} \sim m_b
(\lambda^2,\lambda^2,\lambda^2)$. In order to obtain operators 
$\mathrm{SCET}_{\mathrm{I}}$ from the full theory, we integrate the
modes of order $m_b$, and the Wilson coefficients are determined
by matching the full theory and $\mathrm{SCET}_{\mathrm{I}}$ at $\mu = 
m_b$. 

The typical momentum scale in $\mathrm{SCET}_{\mathrm{I}}$ is $p^2
\sim m_b \Lambda$, which are integrated out to obtain
$\mathrm{SCET}_{\mathrm{II}}$. Here the 
small parameter $\lambda$ scales as $\Lambda/m_b$. The matching
between $\mathrm{SCET}_{\mathrm{I}}$ and $\mathrm{SCET}_{\mathrm{II}}$
is performed at $\mu_0 = \sqrt{m_b \Lambda}$, from which we
obtain jet functions.The dependence of the jet functions on
$1/(m_b\Lambda)$ are unambiguously determined by the matching with
$\mathrm{SCET}_{\mathrm{I}}$ at a given order of the perturbation
theory.   

Let $\xi$ ($\chi$) be the collinear field moving in the
$n^{\mu}$ ($\overline{n}^{\mu}$) direction, satisfying $\FMslash{n}
\xi =0$ ($\FMslash{\overline{n}} \chi =0$). Then the four-quark
operators for nonleptonic $B$ decays in SCET is generically of the
form $(\overline{\xi} \Gamma_1 h)(\overline{\chi} 
\Gamma_2 \chi)$ or $(\overline{\xi} \Gamma_1 \chi)( \overline{\chi}
\Gamma_2 h)$, where $\Gamma_1$, $\Gamma_2$ are Dirac matrices, $h$
is the heavy quark field. The two
types of the operators are obtained from the full-theory
operator $(\overline{q}_1 \Gamma_1 b)(\overline{q}_2 \Gamma_2 q_3)$ by
replacing light quarks by the collinear quarks.
From now on, we consider the operators of the form $(\overline{\xi}
\Gamma_1 h)(\overline{\chi} \Gamma_2 \chi)$. That is,
we shuffle the second type of the operators using Fierz
transformations. In matching we also consider the corresponding
operators with the Fierz transformation in the full theory.
These two operators and their Wilson coefficients are distinct in SCET.

\begin{figure}[b]
\begin{center}
\epsfig{file=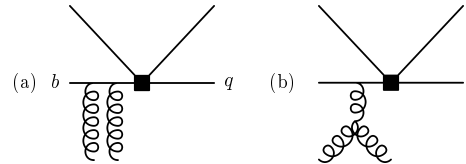, width=6.0cm}
\end{center}
\caption{QCD diagrams attaching two gluons to external fermions to
integrate out off-shell modes. For a heavy quark, the gluons are $A_n$,
$A_{\overline{n}}$ or $A_s$. Diagrams with two gluons attached to
other fermions are omitted.}  
\label{fig1}
\end{figure}

The main issue is to find gauge-invariant four-quark operators by
integrating out off-shell modes when we attach gluons to external fermion
lines. We consider gluon attachment to order $g^2$, as shown in
Fig.~\ref{fig1}. It corresponds to matching the full theory onto
$\mathrm{SCET}_{\mathrm{II}}$, which is equivalent to the two-step
matching since the Wilson coefficients do not depend on the soft
momenta. We employ the two-step matching explicitly for the
time-ordered products. Here we only 
show gluons attached to a heavy quark but it is
straightforward to attach gluons to other fermions. The final result
depends on the following combinations 
\begin{equation}
A=\frac{\overline{n} \cdot A_n}{\overline{n}\cdot q_n}, \ B=
\frac{n\cdot A_{\overline{n}}}{n\cdot q_{\overline{n}}}, \ C=
\frac{\overline{n} \cdot A_s}{\overline{n} \cdot q_s}, \ D=
\frac{n\cdot A_s}{n\cdot q_s},
\end{equation}
where $A_n^{\mu}$ ($A_{\overline{n}}^{\mu}$) are the collinear gauge
field in the $n^{\mu}$ ($\overline{n}^{\mu}$) direction, and
$A_s^{\mu}$ is the soft gauge field in $\mathrm{SCET}_{\mathrm{II}}$,
of order $p_s^2 \sim \Lambda^2$.
And the Wilson lines corresponding to each type of gluons are obtained
by exponentiating the above factors as
\begin{eqnarray}
W&=& \sum_{\mathrm{perm}} \exp \Bigl[ -g\frac{1}{\overline{n} \cdot
    {\cal P}} \overline{n}\cdot A_n \Bigr], \nonumber \\
S&=& \sum_{\mathrm{perm}} \exp \Bigl[ -g \frac{1}{n \cdot
    {\cal R}}n \cdot A_s \Bigr],
\end{eqnarray}
where $\mathcal{P}^{\mu}$ ($\mathcal{Q}^{\mu}$) is the operator
extracting the label momentum in the $n^{\mu}$ ($\overline{n}^{\mu}$)
direction, and $\mathcal{R}^{\mu}$ is the 
operator extracting the soft momentum. The Wilson lines $\overline{W}$
and $\overline{S}$ are obtained from $W$ and $S$ by switching
$n^{\mu}$ and $\overline{n}^{\mu}$.

By integrating out off-shell modes, the singlet and nonsinglet
operators of the form  
$(\overline{\xi}_{\alpha} \Gamma_1 h_{\alpha})(\overline{\chi}_{\beta}
\Gamma_2 \chi_{\beta})$ and $(\overline{\xi}_{\beta} \Gamma_1
h_{\alpha})(\overline{\chi}_{\beta} \Gamma_2 \chi_{\alpha})$ become
gauge-invariant operators, which are given by
\begin{eqnarray}
(\overline{\xi}_{\alpha} \Gamma_1 h_{\alpha})(\overline{\chi}_{\beta}
\Gamma_2 \chi_{\beta})&\rightarrow& H^S_{\alpha
  \beta} L^S_{\gamma\delta} (\overline{\xi}_{\alpha} \Gamma_1 h_{\beta})
(\overline{\chi}_{\gamma} \Gamma_2 \chi_{\delta}), \nonumber \\
(\overline{\xi}_{\beta} \Gamma_1 h_{\alpha})(\overline{\chi}_{\alpha}
\Gamma_2 \chi_{\beta})&\rightarrow& H^N_{\gamma\beta}
L^N_{\alpha\delta} (\overline{\xi}_{\alpha} \Gamma_1 h_{\beta}) 
(\overline{\chi}_{\gamma} \Gamma_2 \chi_{\delta}),
\end{eqnarray}
where $H^O_{\alpha\beta}$, $L^O_{\gamma\delta}$ ($O=S, N$) are color
factors. For singlet operators to order $g^2$, we have
\begin{equation}
H_{\alpha\beta}^S = \Bigl[g(-A+D)-g^2 AD\Bigr]_{\alpha \beta}, \
L_{\delta\gamma}^S = \delta_{\delta\gamma},
\end{equation}
which implies the ordering of the Wilson lines
as $(WS^{\dagger})_{\alpha\beta}$. For nonsinglet operators, we have 
\begin{eqnarray}
H_{\gamma\beta}^N &=& \Bigl[g (-B+C) -g^2 BC \Bigr]_{\gamma\beta},
\nonumber \\
L_{\alpha\delta}^N &=& \Bigl[ g (-A+B-C+D) +g^2 (AC+DB \nonumber \\
&& -CB-DC-AB-AD)\Bigr]_{\alpha\delta},
\end{eqnarray}
which suggests $(\overline{W}
\overline{S}^{\dagger})_{\gamma\beta}$ and $(WS^{\dagger}\overline{S}
\overline{W}^{\dagger})_{\alpha\delta}$. Therefore the gauge-invariant
operators in $\mathrm{SCET}_{\mathrm{II}}$ have the form
\begin{eqnarray}
\label{giop}
O_i^S &=& \Bigl((\overline{\xi} W)_{\alpha} \Gamma_{1i}
(S^{\dagger}h)_{\alpha} \Bigr) \Bigl(
(\overline{\chi}\overline{W})_{\beta} \Gamma_{2i}
(\overline{W}^{\dagger}\chi)_{\beta}\Bigr),  \\ 
O_i^N&=& \Bigl( (\overline{\xi} WS^{\dagger} \overline{S})_{\beta}
\Gamma_{1i} (\overline{S}^{\dagger} h )_{\alpha} \Bigr)
\Bigl( (\overline{\chi} \overline{W})_{\alpha} \Gamma_{2i}
(\overline{W}^{\dagger} \chi)_{\beta} \Bigr), \nonumber
\end{eqnarray}
where $\Gamma_{1i}$, $\Gamma_{2i}$ are appropriate Dirac matrices. 

All the four-quark operators for nonleptonic $B$ decays can be written
in terms of $O_i^S$ and $O_i^N$. And
the Wilson coefficients are determined by matching the full theory
onto SCET. As shown in Eq.~(\ref{giop}) explicitly, the gluon exchange
is allowed only in each current sector, and this holds to all orders in
$\alpha_s$. For example, the collinear gluon $A_{\overline{n}}^{\mu}$
can be exchanged in the light-to-light current sector, but it cannot
interact with the heavy-to-light current. It is true for other gluons
$A_n^{\mu}$ and $A_s^{\mu}$. 
Therefore the matrix elements of these operators can be
reduced to the products of the current matrix elements. This method
was previously referred to as the naive factorization \cite{naive},
which was assumed by the argument of the color transparency
\cite{trans}. But in SCET, this 
factorization property is justified to all orders in $\alpha_s$ for
nonleptonic $B$ decays since the form of the operators
$O_i^{S,N}$ remains the same at higher orders in $\alpha_s$, though
the Wilson coefficients are different.

We have to include spectator interactions because it
is of the same order as the leading contribution from four-quark
operators. The Lagrangian for usoft and collinear quarks in
$\mathrm{SCET}_{\mathrm{I}}$ at order
$\lambda$ is given by $\mathcal{L}_{\xi q}^{(1)}
=\overline{q}_{\mathrm{us}} W^{\dagger} i\FMslash{D}_n^{\perp} \xi +
\mbox{h.c.}$ \cite{beneke1,pirjol}. We consider the time-ordered
products of the four-quark operators including $A_{n\perp}^{\mu}$ with
the effective Lagrangian at subleading order in
$\mathrm{SCET}_{\mathrm{I}}$. When the leading 
four-quark operators $O_i^{S,N}$ do not contribute to the spectator
interactions, we include the subleading operators suppressed by
$\lambda$ in $\mathrm{SCET}_{\mathrm{I}}$. In order to  go down to
$\mathrm{SCET}_{\mathrm{II}}$, we redefine the 
fields by factorizing the usoft modes. We
integrate out the modes of order $p^2 \sim m_b \Lambda$, and 
in matching we obtain jet functions and the dependence of the jet
functions on $1/m_b \Lambda$ is determined in this matching. 

The spectator interactions fall into two
categories. First a collinear gluon from the light-to-light current
can interact with the spectator quark, which we refer to as the
nonfactorizable spectator contributions. Secondly, a collinear gluon
from the heavy-to-light current can interact with the spectator quark,
which contributes to the form factor for the heavy-to-light current.

Including the spectator contributions, the decay amplitudes $A
[\mathcal{O}_i]$ obtained from the four-quark operator $\mathcal{O}_i$
can be written as
\begin{equation}
A [\mathcal{O}_i] = Q_i + N_i + F_i,
\end{equation}
where $Q_i$ are the amplitudes from the four-quark operators, $N_i$
are the nonfactorizable spectator contributions and $F_i$ are the
spectator contributions
to the form factor. We consider each of them in turn, and prove that
each contribution can be written in a factorized form to all orders in
$\alpha_s$.

Let us first consider the contribution of the four-quark operators
themselves. The matrix elements of the four-quark operators are given
as 
\begin{eqnarray}
\label{qi}
Q_i &=& \int d\eta C_{\mathrm{eff},i}^Q (\eta,\mu_0,\mu) \langle
\overline{\xi} W \Gamma_{1i} S^{\dagger} h  \\
&&\times \overline{\chi} \overline{W} \delta (\eta-\mathcal{Q}_+)
\Gamma_{i2} \overline{W}^{\dagger}\chi \rangle \nonumber \\
&=& im_B^2 f_{M2} \zeta (\mu_0,\mu)\int du C_{\mathrm{eff},i}^Q
(u,\mu_0) \phi_{M2} (u,\mu), \nonumber
\end{eqnarray}
where $\mathcal{Q}_+ = n\cdot (\mathcal{Q}^{\dagger}+
\mathcal{Q})$. Here the effective Wilson coefficients 
$C_{\mathrm{eff},i}^Q$ are obtained by matching the full theory onto 
$\mathrm{SCET}_{\mathrm{I}}$, and they are functions of the operators
$\overline{n}\cdot \mathcal{P}/m_b$ and $n\cdot \mathcal{Q}/m_b$,
evaluated at $\mu=m_b$. They are evolved down to
$\mu_0=\sqrt{m_b \Lambda}$ to be matched onto
$\mathrm{SCET}_{\mathrm{II}}$, and again they evolve down to the scale
$\mu$, where the matrix elements are evaluated. 
For example, with $\Gamma_{1i}\otimes \Gamma_{2i} = \gamma_{\mu}
(1-\gamma_5)\otimes \gamma^{\mu} (1 \mp \gamma_5)$, the matrix
elements in Eq.~(\ref{qi}) are given by \cite{bauer2}
\begin{eqnarray}
&&\pm if_{M1} E \int du \, C(u) \phi_{M1} (u) \langle M_2|\overline{\xi}
W\FMslash{\overline{n}} (1-\gamma_5) S^{\dagger}
h|\overline{B}\rangle \nonumber \\
&&= \pm im_B^2 \zeta f_{M1} \int_0^1 du  \, C(u) \phi_{M1} (u), 
\label{mt}
\end{eqnarray}
using the definition of the light-cone wave function \cite{bauer3} 
\begin{eqnarray}
&&\langle M_1| \overline{\chi} \overline{W} \FMslash{n} \gamma_5
  \delta (\eta - \mathcal{Q}_+ )
  \overline{W}^{\dagger} \chi |0\rangle \nonumber \\
&&= -if_{M1} 2E \int_0^1 du \delta [\eta -(4u-2) E]\phi_{M1} (u). 
\end{eqnarray}
And we also use
$\langle M_2|\overline{\xi} W\FMslash{\overline{n}} S^{\dagger}
h|\overline{B}\rangle = 2m_B \zeta$, where $\zeta$
is a nonperturbative function introduced in
Ref.~\cite{charles} ($\xi_P$ in Ref.~\cite{chay}). The heavy-to-light
form factors are reduced to a single universal nonperturbative
function $\zeta$ in SCET.

For the nonfactorizable spectator contributions, we need to consider
the subleading operators in which the lihgt-to-light current contains
$A_n^{\mu}$. There are only two independent operators $O^{(1a,1b)}$ at
subleading order in $\lambda$, which are given by
\begin{eqnarray}
O^{(1a)}_i &=& \Bigl( (\overline{\xi}W)_{\beta} \Gamma_{1i} h_{\alpha} 
  \Bigr) \nonumber \\
&\times&  \Bigl(\overline{\chi} \overline{W} \frac{1}{n\cdot
    \mathcal{Q}^{\dagger}} \frac{\FMslash{n}}{2} \Bigl[ W^{\dagger}
  i \loarrow{\FMSlash{D}}_n^{\perp} W\Bigr]\Bigl)_{\alpha} \Gamma_{2i} 
(\overline{W}^{\dagger} \chi)_{\beta}, \nonumber \\
O^{(1b)}_i &=&\Bigl( (\overline{\xi}W)_{\beta} \Gamma_{1i} h_{\alpha}
  \Bigr) \nonumber \\
&\times& (\overline{\chi} \overline{W})_{\alpha} \Gamma_{2i} \Bigl(
  \Bigl[ W^{\dagger} 
  i\roarrow{\FMSlash{D}}_n^{\perp} W\Bigr] \frac{\FMslash{n}}{2}
  \frac{1}{n \cdot \mathcal{Q}} \overline{W}^{\dagger} \chi
 \Bigr)_{\beta}.
\end{eqnarray}

In $\mathrm{SCET}_{\mathrm{I}}$, the matrix elements of the
time-ordered products for the nonfactorizable
spectator contribution is given as
\begin{equation}
\int d^4 x T\Bigl[ \Bigl( C_i^{(1a)} O^{(1a)}_i +
  C_i^{(1b)}O^{(1b)}_i \Bigr) (0), 
  i\mathcal{L}_{\xi q}^{(1)} (x)   \Bigr],
\label{tor}
\end{equation}
where $C_i^{(1a,1b)}$ are the Wilson coefficients of the operators
$O^{(1a,1b)}$, which are 1 at tree level.
\begin{figure}[h]
\begin{center}
\epsfig{file=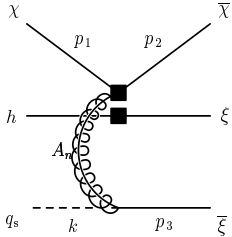, width=3.0cm}
\end{center}
\caption{A Feynman diagram for nonfactorizable spectator
interaction. A subleading operator from
the heavy-to-light current contributes to the form factor. The soft
momentum $k$ is incoming, and $p_i$ ($i=1,2,3,4$) are outgoing.}
\label{fig2}
\end{figure}

The Feynman diagram for the nonfactorizable spectator contribution is
shown in Fig.~\ref{fig2}.  To show the factorization in
$\mathrm{SCET}_{\mathrm{II}}$, we 
decouple the collinear-usoft interaction using the field redefinitions
\cite{bauer4} 
\begin{eqnarray}
&&\xi^{(0)} = Y^{\dagger} \xi, \  A_n^{(0)} = Y^{\dagger} A_n Y,
  \nonumber \\ 
&&  \chi^{(0)} = \overline{Y}^{\dagger} \chi,\ A_{\overline{n}}^{(0)} =
  \overline{Y}^{\dagger} A_{\overline{n}} \overline{Y}, 
  \nonumber \\
&&Y(x) = \mathrm{P} \exp \Bigl( ig \int_{-\infty}^x ds n\cdot
  A_{\mathrm{us}} (ns) \Bigr), \nonumber \\
&&\overline{Y}(x) = \mathrm{P} \exp \Bigl( ig \int_{-\infty}^x ds
  \overline{n} \cdot A_{\mathrm{us}} (\overline{n}s) \Bigr).
\end{eqnarray}
The fields with the superscript 0 are fundamental objects in
$\mathrm{SCET}_{\mathrm{II}}$ and we dropt the subscript from now on. 
We rename the usoft fields in $\mathrm{SCET}_{\mathrm{I}}$ as the
soft fields and we write $Y\rightarrow S$ and $\overline{Y}
\rightarrow \overline{S}$. Then the operators are
matched onto the operators in $\mathrm{SCET}_{\mathrm{II}}$ at  $\mu_0
\sim \sqrt{m_b \Lambda}$.  

The nonfactorizable spectator contribution comes from the matrix
elements of six-quark operators. In calculating the matrix elements,
we first project out color indices in such a way that the 
quark bilinears forming a meson are color singlets. The time-ordered
product of Eq.~(\ref{tor}) in $\mathrm{SCET}_{\mathrm{II}}$ is
proportional to 
\begin{equation}
  \int d\eta d\overline{\eta} dr_+ e^{ir_+
    \overline{n} \cdot x/2} 
J_i^N (\eta, \overline{\eta}, r_+) \mathcal{O}_i
  (\eta,\overline{\eta}, r_+), \nonumber
\label{top2}
\end{equation}
where $J_i^N$ are the jet functions which are obtained in
matching $\mathrm{SCET}_{\mathrm{I}}$ and
$\mathrm{SCET}_{\mathrm{II}}$, and the operator $\mathcal{O}_i$ are
given as 
\begin{eqnarray}
\mathcal{O}_i &=& \Bigl[\overline{\xi} W \Gamma_{1i} \delta (\eta
-\mathcal{P}_+) W^{\dagger} \xi \Bigr] 
\Bigl[ \overline{\chi} \overline{W} \Gamma_{2i} \delta
  (\overline{\eta} - \mathcal{Q}_+) \overline{W}^{\dagger} \chi
  \Bigr]\nonumber \\
&&\times \Bigl[ \overline{q}_s \FMslash{n} (1-\gamma_5) \delta (
  \mathcal{R}^{\dagger} -r_+) S^{\dagger} h \Bigr].
\end{eqnarray}
Since the operator $\mathcal{O}_i$ does not allow gluon exchange between
different sectors, the matrix element can be reduced to the
factorized form in terms of the products of the matrix elements of the
currents. Evaluating the matrix element of $\mathcal{O}_i$, the
nonfactorizable spectator contributions $N_i$ are given as
\begin{eqnarray}
\label{ni}
N_i &=& \int du dv dr_+ C_{\mathrm{eff},i}^N (\mu_0) J_i^N
(u,v,r_+,\mu_0, \mu) \\
&\times& A_i f_B f_{M1} f_{M2} \phi_{M1} (u,\mu) \phi_{M2} (v,\mu)
\phi_B^+ (r_+,\mu), \nonumber
\end{eqnarray}
where $C_{\mathrm{eff},i}^N$ are the Wilson coefficients, and 
$A_i$ are the normalization constants when we evaluate $\langle
\mathcal{O}_i \rangle$. The variables $u$, $v$ are given as
$u=\eta/(4E) +1/2$ and $v=\overline{\eta}/(4E) +1/2$. 
In Eq.~(\ref{ni}), the leading-twist light-cone wave function for the
$B$ meson is given by \cite{beneke2}
\begin{equation}
\Psi_B (r) =-\frac{if_B m_B}{4} \Bigl[\frac{1+\FMslash{v}}{2}
  \Bigl( \FMslash{\overline{n}} \phi_B^+ (r) +\FMslash{n}
  \phi_B^- (r) \Bigr) \gamma_5  \Bigr].
 \end{equation}

\begin{widetext}
At tree level, the matrix element of Eq.~(\ref{tor}) is given by
\begin{eqnarray}
&& - \alpha_s \frac{C_F}{N^2} \int
d\overline{n}\cdot x \int dr_+ e^{i r_+ \overline{n} \cdot x/2}
\nonumber \\
&\times& \Bigglb\{ 
\langle \Bigl[(\overline{\xi} W)_{\beta} \Gamma_{1i} (S^{\dagger}
  h)_{\alpha} \Bigr] 
\Bigl[(\overline{\chi} \overline{W} )_{\gamma} \frac{1}{n\cdot
    \mathcal{Q}^{\dagger}} 
\frac{\FMslash{n}}{2} \gamma_{\perp}^{\mu} \Gamma_{2i}
(\overline{W}^{\dagger} \chi)_{\gamma} \Bigr] \Bigl[
  (\overline{q}_{\mathrm{s}} 
S)_{\alpha} (\overline{n}\cdot x) \frac{1}{n\cdot \mathcal{R}^{\dagger}}
\gamma_{\mu}^{\perp}\frac{1}{\overline{n}\cdot   \mathcal{P}}
  (W^{\dagger} \xi)_{\beta} (0) \Bigr] 
\rangle \nonumber \\
&&+ \langle \Bigl[(\overline{\xi} W)_{\beta}
  \Gamma_{1i} (S^{\dagger}   h)_{\alpha} \Bigr] 
\Bigl[(\overline{\chi} \overline{W} )_{\gamma}
\Gamma_{2i}  \gamma_{\perp}^{\mu} \frac{\FMslash{n}}{2} \frac{1}{n\cdot
  \mathcal{Q}} 
(\overline{W}^{\dagger} \chi)_{\gamma} \Bigr] \Bigl[
  (\overline{q}_{\mathrm{s}} 
S)_{\alpha}(\overline{n}\cdot x)  \frac{1}{n\cdot
  \mathcal{R}^{\dagger}}\gamma_{\mu}^{\perp}
\frac{1}{\overline{n}\cdot 
  \mathcal{P}} (W^{\dagger} \xi)_{\beta} (0)\Bigr]
\rangle
\Biggrb\}. 
\label{spec}
\end{eqnarray}
\end{widetext}

For example, with the Dirac structure $\gamma_{\mu}
(1-\gamma_5) \otimes \gamma^{\mu} (1\pm \gamma_5)$, the matrix element
of Eq.~(\ref{spec}) is given by
\begin{eqnarray}
&&i\frac{C_F}{N^2} \pi\alpha_s
f_B f_{M1} f_{M2} m_B \int dr_+ \frac{\phi_B^+ (r_+)}{r_+}
\nonumber \\ 
&&\times \int du \frac{\phi_{M1} (u)}{u} \int dv \frac{\phi_{M_2}
  (v)}{v}.
\end{eqnarray}
For $(1-\gamma_5)\otimes (1+\gamma_5)$, it is zero at
leading order.

For the spectator contribution to the heavy-to-light form factor, the
subleading operators in $\mathrm{SCET}_{\mathrm{I}}$ are given by 
\begin{eqnarray}
J_i^{(0)} &=& ( \overline{\xi} W \Gamma_{1i} h)(\overline{\chi}
\overline{W} \Gamma_{2i} \overline{W}^{\dagger} \chi),  \\
J_i^{(1a)} &=& \Bigl[ \overline{\xi} W (W^{\dagger}
  i\FMSlash{D}_{n\perp} W) \frac{\Gamma_{1i}}{\overline{n} \cdot
    \mathcal{P}^{\dagger}} h\Bigr] (\overline{\chi} \overline{W} \Gamma_{2i}
\overline{W}^{\dagger}\chi ), \nonumber \\
J_i^{(1b)} &=& \Bigl[ \overline{\xi}W (W^{\dagger}
i\FMSlash{D}_{n\perp} W) \frac{\Gamma_{1i}}{m_b} h\Bigr] 
(\overline{\chi} \overline{W} \Gamma_{2i}
\overline{W}^{\dagger} \chi ). \nonumber  
\end{eqnarray}

\begin{figure}[h]
\begin{center}
\epsfig{file=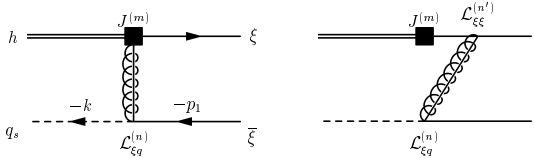, width=7.0cm}
\end{center}
\caption{Tree-level graphs in $\mathrm{SCET}_{\mathrm{I}}$ for the
  spectator contribution to the heavy-to-light form factor. The first
  diagram contributes to $T_{1,2,4}$, and the second diagram
  contributes to $T_{0,1,3,4,5,6}$.}
\label{fig3}
\end{figure}

At leading order in SCET, the relevant factorizable time-ordered
products are given as
\begin{equation}
T_{0i}^F = T[J_i^{(0)}, i\mathcal{L}_{\xi q}^{(1)} ]\equiv \int d^4 x
T[J_i^{(0)} (0) i\mathcal{L}_{\xi q}^{(1)} (x) ], 
\end{equation}
as well as
\begin{eqnarray}
T_{1i}^F &=& T[J_i^{(1a)}, i\mathcal{L}_{\xi q}^{(1)}], \ T_{2i}^F =
T[J_i^{(1b)}, i\mathcal{L}_{\xi q}^{(1)}], \nonumber \\
T_{3i}^F &=& T[J_i^{(0)}, i\mathcal{L}_{\xi q}^{(2b)}], \ T_{4i}^{NF}
= T[J_i^{(0)}, i\mathcal{L}_{\xi q}^{(2a)}],  \\
T_{5i}^{NF}&=& T[J_i^{(0)}, i\mathcal{L}_{\xi\xi}^{(1)},
  i\mathcal{L}_{\xi q}^{(1)}], \ T_{6i}^{NF} =
T[J_i^{(0)},i\mathcal{L}_{cg}^{(1)}, i\mathcal{L}_{\xi
    q}^{(1)}], \nonumber  
\end{eqnarray}
where the effective Lagrangian $\mathcal{L}_{\xi\xi}^{(1)}$,
$\mathcal{L}_{\xi q}^{(2a,2b)}$ and $\mathcal{L}_{cg}^{(1)}$ are
presented in Refs.~\cite{form,pirjol}.These are schematically shown in
Fig.~\ref{fig3}. The factorizable contributions 
$T^F$ are the hard-scattering contributions, and 
the nonfactorizable time-ordered products $T^{NF}$ represent genuine
long-distance physics and we absorb these into soft nonperturbative
parameter $\zeta$. The factorizable time-ordered products in
$\mathrm{SCET}_{\mathrm{II}}$ can be written as
\begin{eqnarray}
\label{fi}
F_i &=& \int du dv dr_+ C_{\mathrm{eff},i}^F (\mu_0), J_i^F
(u,v,r_+,\mu_0,\mu) \\
&\times& A_i f_B f_{M1} f_{M2} \phi_{M1} (u,\mu) \phi_{M2} (v,\mu)
\phi_B^+ (r_+,\mu), \nonumber
\end{eqnarray}
where $C_{\mathrm{eff},i}^F$ are the Wilson coefficients obtained at
$\mu=m_b$ and evovled down to $\mu_0$, and $J_i^F$ are the jet
functions.

If we include all the contributions of the four-quark operators and
the nonfactorizable spectator interactions, the decay amplitudes for
$\overline{B} \rightarrow \pi\pi$ can be obtained at leading order in
SCET. The decay amplitudes involve the matrix elements of four-quark
and six-quark operators, which reduce to the products of the current
matrix elements. This is because the gluon exchange between different
sectors of the currents are not allowed and it holds true to all
orders. The gauge-invariant quark operators remain the same at higher
orders in $\alpha_s$, while only the Wilson coefficients
change. Furthermore, the factorizable contributions are all written as
the convolution integral of the hard-scattering amplitudes with the
meson wave functions, which are finite. Therefore we have proved that
the leading-order contribution to $B$ decays into two light mesons in
SCET is factorized to all orders in $\alpha_s$.

Combining all the contributions, the decay amplitudes to order
$\alpha_s$ are consistent
with the decay amplitudes obtained by Beneke et al. \cite{beneke3}. 
However, the general expression in Eqs.~(\ref{qi}), (\ref{ni}) and
(\ref{fi}) for the decay amplitudes in SCET goes
further than the lowest-order result in Ref.~\cite{beneke3}. The form
of the operators and the factorization properties hold true to all
orders in $\alpha_s$ and they
include the jet functions. Since there are
two scales $m_b$ and $\mu_0$ are involved, the decay amplitudes 
at higher orders in $\alpha_s$ will be different
from the approach in the heavy quark limit.
It remains to be seen whether the factorization properties can be
sustained at subleading order in SCET. However our result is a first
step in understanding nonleptonic $B$ decays in SCET.

This work is supported by  R01-2002-000-00291-0 from the Basic
Research Program of KOSEF.

\end{document}